\documentclass[sigconf, nonacm]{acmart}

\AtBeginDocument{%
  }

\usepackage{algorithm, algorithmic}
\usepackage{array}
\usepackage{multirow}
\usepackage{booktabs}
\usepackage{mathtools} 
\usepackage{enumitem}
\usepackage{subcaption}
\usepackage{pdfpages}
\usepackage{comment}

\begin{document}


\title[Deep Learning for Options Trading: An End-To-End Approach]{Deep Learning for Options Trading:\\ An End-To-End Approach}

\author{Wee Ling Tan}
\authornote{Corresponding Author}
\email{weeling@robots.ox.ac.uk}
\orcid{0000-0003-2664-8238}
\affiliation{%
  \institution{Oxford-Man Institute of Quantitative Finance}
  \institution{University of Oxford}
  \city{Oxford}
  \country{United Kingdom}
}

\author{Stephen Roberts}
\email{sjrob@robots.ox.ac.uk}
\affiliation{%
  \institution{Oxford-Man Institute of Quantitative Finance}
  \institution{University of Oxford}
  \city{Oxford}
  \country{United Kingdom}
}

\author{Stefan Zohren}
\email{stefan.zohren@eng.ox.ac.uk}
\affiliation{%
  \institution{Oxford-Man Institute of Quantitative Finance}
  \institution{University of Oxford}
  \city{Oxford}
  \country{United Kingdom}
}


\begin{abstract}
We introduce a novel approach to options trading strategies using a highly scalable and data-driven machine learning algorithm. In contrast to traditional approaches that often require specifications of underlying market dynamics or assumptions on an option pricing model, our models depart fundamentally from the need for these prerequisites, directly learning non-trivial mappings from market data to optimal trading signals. Backtesting on more than a decade of option contracts for equities listed on the S\&P 100, we demonstrate that deep learning models trained according to our end-to-end approach exhibit significant improvements in risk-adjusted performance over existing rules-based trading strategies. We find that incorporating turnover regularization into the models leads to further performance enhancements at prohibitively high levels of transaction costs.
\end{abstract}



\maketitle

\section{Introduction}
\label{introduction}

Options form a class of derivatives known as contingent claims, granting one of the counterparties the right to transact an underlying asset at a certain time in the future and at a specific price. For instance, the buyer of a call (put) stock option pays a premium to the seller for the right to buy (sell) the underlying stock by a specified expiration date and strike price. Unlike in cash markets that typically exhibit linear payoffs, the ability to generate synthetic and highly non-linear exposures has made options an increasingly important tool among investors as trading instruments alongside other derivatives \cite{black1975fact}.

The options market has continued to grow significantly over the last decade. According to the Options Clearing Corporation, the average daily combined volume of US equity and non-equity options has steadily risen and roughly tripled over the past decade from about 16.3 million in 2013 to 44.2 million contracts in 2023 \cite{occstatistics}. Given the increasing popularity of options trading by market participants conducted in modern electronic exchanges, there exists an unprecedented opportunity to engage in large-scale data analysis of options trading from the perspective of an active investor.

Classical option pricing models and their parametric variants, stemming from the seminal works of Black and Scholes \cite{black1973pricing} and Merton \cite{merton1973theory}, regard options as redundant assets. While these approaches offer a tractable no-arbitrage pricing framework, they necessitate thorough specifications of underlying dynamics and require assumptions such as a frictionless market and the ability of market makers to perfectly hedge their exposures. Practically, these models are often vulnerable to model misspecification and are usually too simplistic to sufficiently account for empirically observed variations in option returns \cite{buchner2022factor}. Recent works have challenged this framework, demonstrating that option prices are influenced by additional risks beyond the underlying asset's exposure \cite{coval2001expected, goyal2009cross, heston2023option}, with several studies documenting the existence of mispricing in the options market \cite{bali2023option, eisdorfer2022maturity}.

Despite the accelerating growth of the options market and the evidence of options mispricing, there is a surprising lack of research on a scalable machine learning-based strategy that is able to directly leverage the abundance of data to conduct options trading on behalf of an active investor. In this work, we address this gap by introducing a novel class of deep learning models capable of managing and trading a portfolio of options in a highly data-driven approach. 

Previous research has conventionally approached the complexities of managing portfolios of options by developing methods for optimal hedging and accurately pricing options. Our model departs fundamentally from the need for these prerequisites, and are directly optimized to learn highly non-trivial mappings from observed data to optimal trading decisions based on risk-adjusted performance. The resulting end-to-end framework does not depend on specific market dynamics, and can be extended broadly across instruments where market data is available.

To illustrate our approach, we construct the framework by training end-to-end neural networks of different complexities and show that our models are able to outperform existing rules-based strategies in managing a portfolio of options, demonstrating strong risk-adjusted returns over a backtest period of over a decade. We subsequently extend our models to incorporate turnover regularization during optimization, leading to further performance enhancements in the presence of high transaction costs.

\section{Related Work}

To place our work within the context of existing research on options, we briefly review the broad literature in three areas: replication and hedging, pricing and valuation, and predicting option returns. We subsequently demonstrate how our work contributes to existing systematic trading strategies within options markets.

\subsection{Options Literature}

Traditionally, options have been extensively studied due to their significance in both replication and hedging strategies. The works of \cite{buehler2019deep} and \cite{kolm2019dynamic} apply reinforcement learning techniques towards approximating policies for optimal hedging in the presence of trading costs. While these works are pertinent to a product or liquidity provider needing to accurately hedge its exposures to books of options and other complex derivatives, our analysis fundamentally differs in its focus on options trading from the standpoint of an active investor seeking to profit from an options trading strategy and whose main objective is not predominantly about neutralizing exposures. Furthermore, hedging techniques which are based on reinforcement learning often require generating or simulating possible market paths to arrive at an optimal trading strategy. In contrast, the solution we offer does not require any assumption or simulation of market processes, and scales with available historical data and compute.

There has also been significant emphasis placed on the need to accurately price or value an option in order to facilitate optimal hedging and trading, with notable contributions from the classical Black-Scholes-Merton option pricing model for European-style options \cite{black1973pricing, merton1973theory}. Non-parametric models have also been developed, with \cite{hutchinson1994nonparametric} and \cite{ivașcu2021option} using neural networks to approximate the market's option pricing function. Conversely, our framework focuses on automatically extracting features and making trading decisions directly from available data, effectively circumventing the need for engineering an option pricing or valuation model.

Several studies have also employed machine learning for predicting option returns, often framing the trading strategy as a standard regression problem and subsequently determining the direction of price movements based on forecasted returns. \citet{bali2023option} use both linear and nonlinear models and ex-ante option-based and stock-based characteristics to predict monthly returns of delta-hedged options, demonstrating high out-of-sample forecast accuracy. However in the case of a trend-following strategy, \cite{lim2019enhancing} shows that models that accurately predict returns do not necessarily guarantee a positive or superior strategy performance, given that the overall profitability of trading strategies is influenced by other factors including the distribution of returns, position sizes and the presence of risk adjustments like volatility targeting \cite{harvey2018impact}. Taking this into consideration, our work effectively integrates both trend prediction and optimal position sizing simultaneously within a single end-to-end function, eliminating the need to forecast option returns in making trading decisions.

\subsection{Systematic Strategies and Risk Premia in Options Markets}

Our work contributes to a series of research documenting systematic strategies originating from studies demonstrating additional risk factors \cite{buchner2022factor, coval2001expected, goyal2009cross, heston2023option, vasquez2017equity}, and the existence of mispricing \cite{bali2023option, eisdorfer2022maturity} within the options market. Considering the broad range of possible risk factors and strategies, we concentrate our analysis on systematic trend-based strategies. Given their simplicity of being based primarily on historical price trends and returns, and unlike volatility-based strategies, trend-based strategies often do not require additional assumptions on any underlying option pricing model or specifications of market dynamics.

\subsection{Momentum, Mean-Reversion and Applications of Machine Learning}
Trend-based strategies consists of primarily momentum and mean-reversion strategies.
Momentum strategies operate on the principle that asset returns exhibit a tendency to persist in the same direction and aim to trade in the direction of the trend \cite{jegadeesh1993returns, moskowitz2012time}. On the other hand, mean-reversion strategies adopt an opposite and contrarian view, taking on positions that bet on the eventual break down and correction of overextended trends \cite{de1985does, poterba1988mean}. Recently, machine learning models have been increasingly utilized in trend-based trading strategies \cite{lim2019enhancing, poh2021building, poh2022enhancing, tan2023spatio, wood2021trading, wood2022slow}.

While momentum strategies have been extensively documented in a range of asset classes from cash equities \cite{rouwenhorst1998international} to futures markets \cite{moskowitz2012time}, these strategies have received relatively less attention in options markets. \citet{heston2023option} document strong evidence of the presence of momentum in options markets and propose a series of rules-based momentum strategies. However, these rules-based strategies often require explicit specifications for the trading rule, often with insufficient evidence to justify selecting a certain rule over another. In light of these observations, we employ deep neural networks to automatically learn risk-adjusted trading rules in a data-driven approach, and provide a thorough comparison between these trend-based strategies.

\section{Overview of Dataset}

Our dataset consists of option contracts sourced from the OptionMetrics Ivy DB database, comprising end-of-day bid-ask prices, implied volatility and Greeks of individual options. We independently verify the accuracy of the provided implied volatility and Greeks using a binomial tree model of  \citet{cox1979option} only for the purpose of initial delta hedging (refer to Section \ref{option_strategies}), but otherwise we do not assume any underlying option pricing model throughout this work. We focus our analysis on option contracts of equities listed on the S\&P 100 Index, as these companies span a wide range of sectors, representing major large-cap optionable companies in the US market, and are associated with higher option liquidity. We obtain underlying stock prices from the Center for Research in Security Prices (CRSP), which we use for computing option moneyness and accounting for corporate actions such as stock splits. We perform backtesting with the most recent market data (as of this work) from 2010 to 2023, which notably includes the market selloff following the COVID-19 pandemic. 

We impose a series of data filters established in the literature to ensure the consistency of our analysis. We consider only standard monthly option contracts that expire on the third Friday of the month and exclude any special settlements due to corporate actions. We then exclude options that contain price observations that breach American-style option bounds, having a bid price of zero, or where the ask price is smaller or equal to the bid price, and require options to have a positive open interest on the day of portfolio formation. In addition, we disregard options that contain one or more missing observations between the day of portfolio formation and expiration.

On the expiration day of each month and for each stock, we form portfolios of static delta-neutral straddle options by selecting a pair of call and put contracts with identical strike prices that are closest to at-the-money (ATM) and expire in the following month. Since it is usually not possible to select options with moneyness exactly equal to 1.0 (moneyness of call = $S/K$, put = $K/S$ where $K = $ strike price, $S = $ stock price), we select the pair of options that are closest to ATM within a moneyness range of 0.95 to 1.05. 

Our resulting universe consists of 29984 options that are traded, with a total of 603068 daily returns observations throughout the backtest period. The returns of straddle options have positive means (1.41\% monthly), large standard deviations (90.85\% monthly), and significant positive skewness as indicated by a low median (-15.61\% monthly). We provide an in-depth explanation on portfolio formation and returns computation in Section \ref{option_strategies}.

\section{Systematic Options Trading Strategies}
\label{option_strategies}
Let $i = 1, 2, \cdots, N_t$ denote individual underlying stocks. For a given portfolio of (1-month, ATM, static delta-neutral) straddle options of these stocks that is rebalanced daily, the overall returns of a strategy that equally diversifies over $N_t$ straddles at day $t$ can be expressed as follows:
\begin{align}
    r_{t, t+1}^{\text{STRATEGY}} &= \frac{1}{N_t} \sum_{i=1}^{N_t} X_t^{(i, \text{straddle})}~\left( \frac{\sigma_{\text{tgt}}}{\sigma_t^{(i, \text{straddle})}}\right ) ~r_{t, t+1}^{(i, \text{straddle})} \label{eqn:strategy}
\end{align}
whereby
\begin{align}
    r_{t, t+1}^{(i, \text{straddle})} &= \frac{p_{t+1}^{(i, \text{straddle})} - p_t^{(i, \text{straddle})}}{p_{t}^{(i, \text{straddle})}} \label{eqn:delta_neutral_straddle_returns} \\
    p_t^{(i, \text{straddle})} &= w_{\text{norm}}^{(i, \text{call})} p_t^{(i, \text{call})} + w_{\text{norm}}^{(i, \text{put})} p_t^{(i, \text{put})} \nonumber \\
    w_{\text{norm}}^{(i, \text{call})} &= \frac{w^{(i, \text{call})}}{w^{(i, \text{call})} + w^{(i, \text{put})}} \nonumber \\
    w_{\text{norm}}^{(i, \text{put})} &= 1 - w_{\text{norm}}^{(i, \text{call})} \nonumber \\
    w^{(i, \text{call})} &= - \Delta_0^{(i, \text{put})} \nonumber \\ 
    w^{(i, \text{put})} &= \Delta_0^{(i, \text{call})} \nonumber
\end{align}
On the day of portfolio formation ($t=0$), we construct static delta-neutral straddle options by holding the call and put options with weights $- \Delta_0^{(i, \text{put})}$ and $\Delta_0^{(i, \text{call})}$ respectively, with $ \Delta_0$ representing initial deltas. We subsequently normalize both weights to sum to one, resulting in a respective weightage of the call and put options that is generally close to 50-50. In evaluating the price of the straddle, we take $p_t^{(i, \text{call})}$ and $p_t^{(i, \text{put})}$ to be the bid-ask midpoints of the call and put options.

Consistent with other works \cite{goyal2009cross, heston2023option}, we focus on static delta-neutral straddles as these instruments are on average invariant to movements in the underlying. \cite{tian2023limits} demonstrate that performing one-time delta hedging at initiation neutralizes the overall directional risks associated with an option by about 70\%. As such, we opt for delta hedging at initiation and following similar works, we disregard the early exercise premium embedded in the American-style options and therefore ignore the possibility of early assignment of short options. 

Here, $X_t^{(i, \text{straddle})} \in [-1, 1]$ denotes the trading signal or position for the straddle option of stock $i$ at day $t$, and $r_{t, t+1}^{(i, \text{straddle})}$ is the realized returns of the straddle from $t$ to $t+1$. We find evidence that straddle options in the cross-section of individual constituents of the S\&P 100 exhibit different levels of volatility using a Levene's test at a significance level of 1\%. Given these differences, we include volatility targeting at the level of individual straddle options to scale the realized returns $r_{t, t+1}^{(i, \text{straddle})}$ by their volatility in order to target equal assignments of risk. We set the annualized volatility target $\sigma_{\text{tgt}}$ to be $15\%$ and estimate the ex-ante volatility $\sigma_t^{(i, \text{straddle})}$ with a 20-day exponentially weighted moving standard deviation of daily straddle returns.

All of the following trend-based benchmarks which we incorporate in our work adhere to this general framework and are concerned with constructing an accurate trading signal $X_t^{(i, \text{straddle})}$:

\paragraph{\bf Long Only (Short Only)} This strategy takes a maximum long (short) position $X_t^{(i, \text{straddle})} = 1$ or $-1$ for all straddle options in the portfolio. Since the performance of a short only strategy is the exact opposite of a long only strategy, we focus only on the long only strategy. 

\paragraph{\bf TSMOM (TSMR)} Following the time-series momentum strategy (TSMOM) of \citet{moskowitz2012time}, we adopt the strategy with a monthly horizon. The position taken for a straddle option is based on the sign of the option's returns over the past 20 days: $X_t^{(i, \text{straddle})} = \text{sgn}(r_{t-20, t}^{(i, \text{straddle})})$. In the time-series mean reversion strategy (TSMR), we modify TSMOM to take on a negative loading on past returns, where $X_t^{(i, \text{straddle})} = -\text{sgn}(r_{t-20, t}^{(i, \text{straddle})})$. This strategy takes a contrarian approach by taking on a long (short) position for straddles with negative (positive) returns over the past 20 days.
\paragraph{\bf MACD (MACDMR)} We use volatility normalised moving average convergence divergence (MACD) indicators based on \citet{baz2015dissecting} in place of the sign of returns for estimating the trading signal: 
\begin{align}
    &\text{MACD}(i, t, S, L) = m(i, t, S) - m(i, t, L) \nonumber \\
    &\text{MACD}_{\text{norm}}(i, t, S, L) = \frac{\text{MACD}(i, t, S, L)}{\text{std}(p_{t-5 : t})} \nonumber \\
    &Y^{(i, \text{straddle})}_t = \frac{\text{MACD}_{\text{norm}}(i, t, S, L)}{\text{std}(\text{MACD}_{\text{norm}}(i, t-20 : t, S, L))} \nonumber \\
    &X_t^{(i, \text{straddle})} = \frac{1}{3} \sum_{k=1}^3 \phi(Y_t^{(i, \text{straddle})} (S_k, L_k)) \label{eqn:macd}
\end{align}
where $\text{MACD}(i, t, S, L)$ denotes the MACD value of the straddle option of stock $i$ at day $t$ with a short time scale $S$ and long time scale $L$. $m(i, t, j)$ is defined as the exponentially weighted moving average of the straddle's prices at day $t$, with a time scale $j$ corresponding to a half-life of $HL = \log(0.5) / \log(1 - 1/j)$. We combine MACD signals in an equally weighted sum over multiple short and long time scales $S_{k} \in \{2, 4, 8\}$, $L_{k} \in \{8, 16, 32\}$ to construct a position $X_t^{(i, \text{straddle})}$ where $\phi(y) = \frac{y\exp(\frac{-y^2}{4})}{0.89}$. Likewise, we consider both the momentum $X_t^{(i, \text{straddle})} $ (MACD) and mean reversion $-X_t^{(i, \text{straddle})}$ (MACDMR) strategies. We refer the reader to \cite{baz2015dissecting} for further details on the strategy implementation.

\paragraph{\bf TSHestonMOM (TSHestonMR)} Following \citet{heston2023option}, we consider a strategy that constructs the trading signal for the straddle option of stock $i$ on the day of portfolio formation ($t=0$), holding the position unchanged to expiry: 
\begin{align}
    Y_{0:t}^{(i, \text{straddle})} &= r_{0-n\text{M}, 0}^{(i, \text{straddle})} \label{raw_heston_momentum} \\
    X_{0:t}^{(i, \text{straddle})} &= \text{sgn}(Y_{0:t}^{(i, \text{straddle})})
\end{align}
where $r_{0-n\text{M}, 0}^{(i, \text{straddle})}$ is the average returns for the series of (1-month, ATM, static delta-neutral) straddle options for stock $i$ over a lookback period of $n$ months from the day of portfolio formation. 
A key distinction between TSMOM and TSHestonMOM is that the momentum signals identified in the latter strategy are more accurately referred to as cross-serial correlations, in that the options used to compute the $n$-period average returns $r_{0-n\text{M}, 0}^{(i, \text{straddle})}$ are different from those used to construct the trading signal for the options to be traded following $t=0$. In other words for the TSHestonMOM, all types of momentum signals involve past returns of a set of options predicting the future returns on an entirely new set. Similar to \cite{heston2023option}, we consider multiple strategies corresponding to average returns of monthly straddle returns over various lookback periods ranging from $n = 1, 3, 6, 12$ corresponding to monthly, quarterly, semiannual and annual returns. We again consider both the momentum $X_{0:t}^{(i, \text{straddle})} $ (TSHestonMOM) and mean reversion $-X_{0:t}^{(i, \text{straddle})}$ (TSHestonMR) strategies.

\paragraph{\bf CSHestonMOM (CSHestonMR)} Based on the long-short approach of \cite{heston2023option} and \cite{jegadeesh1993returns}, we implement a cross-sectional momentum strategy that scores and ranks a stock based on its average straddle returns computed as per TSHestonMOM. On the day of portfolio formation, the strategy utilizes a high-minus-low decile portfolio, taking a maximum long and short position for the top and bottom $10\%$ of ranked stocks respectively and holding the positions to expiry. We first calculate raw momentum scores $Y_{0:t}^{(i, \text{straddle})} = r_{0-n\text{M}, 0}^{(i, \text{straddle})}$ according to Equation (\ref{raw_heston_momentum}). Then, we rank stocks by their raw momentum scores:
\begin{align}
    X_{0:t}^{(i, \text{straddle})} &= \{+1, -1, 0\}
\end{align}
where $X_{0:t}^{(i, \text{straddle})} = +1$ for stocks ranked in the top 10\% and $-1$ for the bottom 10\% ranked stocks, and $0$ otherwise. We examine both the momentum $X_{0:t}^{(i, \text{straddle})} $ (CSHestonMOM) and mean reversion $-X_{0:t}^{(i, \text{straddle})}$ (CSHestonMR) strategies.

\section{Deep Learning for Options Trading}

\subsection{General End-To-End Framework}
We frame the problem of generating optimal trading decisions $X_t^{(i, \text{straddle})}$ for a portfolio of options with a model $f$ as an end-to-end framework. Given time $t$ and a straddle option of stock $i$, we have an input $\mathbf{u}_t^{(i, \text{straddle})} \in \mathbb{R}^{d}$ of option features. We learn a model $f$ with trainable parameters $\boldsymbol{\theta}$:
\begin{equation}
\label{eqn:input_to_output}
X_t^{(i, \text{straddle})} = f(\mathbf{u}_t^{(i, \text{straddle})} ; \boldsymbol{\theta})
\end{equation}

In this framework, trading signals $X_t^{(i, \text{straddle})}$ are directly computed using the point-in-time snapshot of the option's features, integrating both trend prediction and optimal position sizing within a single function $f$. Unlike standard supervised learning paradigms, a distinctive characteristic of our end-to-end framework is the unavailability of ground-truth labels for the optimal trading signal $X_t^{(i, \text{straddle})}$ of a given option at any point in time. This necessitates the learning of a non-trivial mapping from an option's features to optimal trading signals. We discuss the choice of architectures and optimization of these models in the following sections.

\subsection{Network Architectures}
Given the problem of learning a mapping from option features to trading signals, it is not immediately clear which choice of architecture would best suit an end-to-end model, warranting the need to consider multiple options. Taking this into account, we examine various choices of neural networks for $f$.

\paragraph{\bf Linear} Beginning with the elementary case of a neural network comprising of a single fully connected layer:
\begin{equation}
\label{eqn:linear}
X_t^{(i, \text{straddle})} 
= g(\mathbf{W}^\top \mathbf{u}_t^{(i, \text{straddle})} + b)
\end{equation}
where $\mathbf{W} \in \mathbb{R}^{m}$, $\mathbf{u}_t^{(i, \text{straddle})} \coloneqq \mathbf{u}_{t-\tau+1 : t}^{(i, \text{straddle})}  \in \mathbb{R}^{m} $ with $m = \tau~d$, $b \in \mathbb{R}$ and $g = \tanh$ is the activation function. The model computes a linear combination of the input features prior to the $\tanh$ activation function that outputs the trading signal to be within the finite range of $[-1, 1]$. To factor in the immediate temporal history of observations for making predictions, we concatenate features from the past $\tau = 5$ days from time $t$ into a single input vector. 

\paragraph{\bf Multilayer Perceptron (MLP)} We include an additional hidden layer to the Linear model, enhancing the model's depth: 
\begin{equation}
\label{eqn:mlp}
X_t^{(i, \text{straddle})} 
= g \lbrack \mathbf{W}^{\lbrack 2 \rbrack\top} \sigma (\mathbf{W}^{\lbrack 1 \rbrack\top}\mathbf{u}_t^{(i, \text{straddle})} + b^{\lbrack 1 \rbrack}) + b^{\lbrack 2 \rbrack} \rbrack
\end{equation}
with $g = \sigma = \tanh$ corresponding to the activation functions of each layer.

\paragraph{\bf Convolutional Neural Networks (CNN)} Modified for time series data, CNNs have been designed to incorporate causal convolutions that utilize only past information for forecasting \cite{binkowski2018autoregressive}, maintaining the autoregressive ordering of temporal features. We consider a 1-D autoregressive CNN:
\begin{align}
\mathbf{h}_t^{(i, \text{straddle})} 
&= P \sigma \lbrack \mathbf{W}_c^{\lbrack 2 \rbrack}\ast \sigma(\mathbf{W}_c^{\lbrack 1 \rbrack}\ast\mathbf{u}_t^{(i, \text{straddle})} + \mathbf{b}_c^{\lbrack 1 \rbrack}) + \mathbf{b}_c^{\lbrack 2 \rbrack} \rbrack \nonumber \\
X_t^{(i, \text{straddle})} 
&= g \lbrack \mathbf{W}^{\lbrack 2 \rbrack\top} \sigma (\mathbf{W}^{\lbrack 1 \rbrack\top}\mathbf{h}_t^{(i, \text{straddle})} + b^{\lbrack 1 \rbrack}) + b^{\lbrack 2 \rbrack} \rbrack
\end{align}
where $\mathbf{W}_c^{\lbrack l \rbrack}$ represent convolutional kernels with bias terms $\mathbf{b}_c^{\lbrack l \rbrack}$ and activation functions $g = \sigma = \tanh$, and $\ast$ represents the causal convolution operator. Subsequently, we perform average pooling with $P$ prior to passing the activations $\mathbf{h}_t$ to a fully connected neural network.

\paragraph{\bf Long Short-term Memory (LSTM)} Recurrent neural networks (RNNs) have traditionally found applications in sequence modelling and time series forecasting \cite{lim2021time}. Given the sequential nature of our prediction task, it is natural to consider recurrent architectures. We implement a single layer LSTM model \cite{hochreiter1997long} that takes in an input sequence of option features $\mathbf{u}_t^{(i, \text{straddle})} \coloneqq \mathbf{u}_{t-\tau+1 : t}^{(i, \text{straddle})}  \in \mathbb{R}^{m} $ with $m = \tau~d$ where $\tau$ represents the length of a trajectory, and subdivide the time series into trajectories of $\tau = 20$ during backpropagation. We omit the technical equations for the LSTM architecture for brevity. 

\subsection{Training Details}
\subsubsection{Loss Function}

To facilitate the learning of a non-trivial mapping from option features to optimal trading signals that effectively balances both risk and reward, we directly calibrate the models using the Sharpe ratio \cite{sharpe1998sharpe}, a risk-adjusted performance metric. Given a set of contemporaneous option features and their respective trading signals $\mathcal{D}_{\Omega} = \{(\mathbf{u}_t^{(i, \text{straddle})},~ X_t^{(i, \text{straddle})} = f(\mathbf{u}_t^{(i, \text{straddle})} ; \boldsymbol{\theta})) \}$ with $\Omega = \{ (i, ~ t) \mid i = 1, \cdots, N^t, ~ t = 1, \cdots, T \}$ denoting all straddle-time pairs, we define the loss $\mathcal{L}_{\text{sharpe}}(\boldsymbol{\theta})$ over $\mathcal{D}_{\Omega}$ as the annualized Sharpe ratio: 
\begin{align}
\mathcal{L}_{\text{sharpe}}(\boldsymbol{\theta}) &= - \frac{\frac{1}{\lvert \Omega \rvert}\sum_{\Omega} R_i(t) \times \sqrt{252}}{\sqrt{\frac{1}{\lvert \Omega \rvert}\sum_{\Omega} R_i(t)^2 - \left[\frac{1}{\lvert \Omega \rvert}\sum_{\Omega} R_i(t)\right]^2}} \\[2.5mm]
R_i(t) &= X_t^{(i, \text{straddle})}~\left(\frac{\sigma_{\text{tgt}}}{\sigma_t^{(i, \text{straddle})}}\right)~r_{t,t+1}^{(i, \text{straddle})}
\end{align}

\subsubsection{Optimization}
Within each in-sample window, we perform a train-validation split with the earlier $90\%$ of data used for calibrating the models and the most recent $ 10\%$ reserved for validation. To calibrate the models, we perform backpropagation using minibatch stochastic gradient descent with Adam \cite{kingma2014adam}, and trigger early stopping with a patience of 25 epochs based on the validation loss. In order to select optimal candidates for each machine learning model, we conduct hyperparameter optimization with 100 iterations of random search. We refer the reader to Appendix \ref{appendix_a} for the detailed description of hyperparameter search ranges. Model calibration was performed on a server equipped with an AMD EPYC7713 CPU and multiple NVIDIA L40 GPUs.

\section{Performance Evaluation}

\subsection{Backtest Details}
Following an expanding window approach, we train all models with every block of 5 additional years. In each block, we fix the weights and hyperparameters of the trained models and evaluate the models out-of-sample in the following 5-year window. We perform model calibration over multiple seeded runs and present the aggregated out-of-sample results in Section \ref{results_and_discussion}.

\subsection{Option Features}

To construct an input of option features $\mathbf{u}_t^{(i, \text{straddle})} \in \mathbb{R}^{d}$ as described in Equation (\ref{eqn:input_to_output}), we include a combination of the predictors used in the strategies outlined in Section \ref{option_strategies}:
\begin{enumerate}[label = \textbf{\Roman*.}]
    \item \textbf{Normalized Returns} -- we use $r^{(i, \text{straddle})}_{t-k, t} \slash (\sigma^{(i, \text{straddle})}_t \sqrt{k})$, representing straddle returns normalized by daily volatility estimates scaled to a time scale $k \in \{1, 5, 10, 15, 20\}$, which corresponds to daily, weekly, biweekly, triweekly and monthly returns. 
    \item \textbf{MACD Indicators} -- we take volatility normalised MACD signals $Y_t^{(i, \text{straddle})} (S_k, L_k)$ from Equation (\ref{eqn:macd}) with short and long time scales $S_{k} \in \{2, 4, 8\}$ and $L_{k} \in \{8, 16, 32\}$.
  
    \item \textbf{Option Momentum Features} -- we expand our set of predictors to include the momentum features as defined in Equation (\ref{raw_heston_momentum}), taking the average returns of straddle options for the stock over lookback periods of $n = 1, 3, 6, 12$ months.
    
    \item \textbf{Core Option Features} -- to facilitate comparability with the trend-based benchmarks outlined in Section \ref{option_strategies}, and to focus on the predictive power of the features above, we maintain a parsimonious set of core option features which includes the log-moneyness (of both call and put options forming the straddle) and days to expiry (DTE, in years). Given that the moneyness and DTE of a straddle option changes over time, these core contract features are necessary in identifying the option at particular stages of its lifespan. Distinctively, we exclude other features such as option implied volatility or sensitivity measures such as Greeks which would require making an assumption of an underlying option pricing model such as the Black-Scholes or the binomial model. Given our focus on delta-neutral straddles, we also exclude underlying stock characteristics and stock returns from our core set of features.
\end{enumerate}

\subsection{Results and Discussion}
\label{results_and_discussion}

\begin{table*}[htbp]
\centering
\caption{Performance Metrics -- Raw Signal Outputs}
\label{table:raw_signals}
\resizebox{\textwidth}{!}{
\begin{tabular}{llllllllll}
\hline \toprule
& \textbf{E[Return]} & \textbf{Vol.} & \textbf{\begin{tabular}[c]{@{}l@{}} Downside \\ Deviation \end{tabular}} & \textbf{MDD} & \textbf{Sharpe} & \textbf{Sortino} & \textbf{Calmar} & \textbf{\begin{tabular}[c]{@{}l@{}} Hit \\ Rate \end{tabular}} & \textbf{$\mathbf{\frac{\text{Ave. P}}{\text{Ave. L}}}$} \\ 
\midrule
{\underline{\textbf{Benchmarks}}} &   &   &   &   &   &   &   &   &  \\
Long Only                 &                   0.055 &                  0.103 &               0.058 &             0.188 &         0.534 &          0.940 &         0.291 &     0.451 &      1.341                                                    \\
TSMOM             & -0.076 &                 0.091 &              0.077 &            0.504 &        -0.827 &         -0.982 &        -0.150 &    0.553 &     0.693                                                    \\
TSMR                      &                   \textbf{0.057*} &                  0.088 &               0.049 &             0.168 &         0.646 &          1.160 &         0.340 &     0.445 &      \textbf{1.406*}                                                    \\ 
MACD                      & -0.038 &                 0.056 &               0.047 &            0.308 &        -0.671 &         -0.800 &        -0.123 &    0.559 &     0.697                                                    \\ 
MACDMR                      &                   0.029 &                  0.054 &               0.030 &             0.106 &         0.529 &          0.945 &         0.271 &     0.443 &      1.388                                                    \\ 
TSHestonMOM                     & 0.002 &                 0.053 &               0.035 &            0.150 &       0.043 &        0.065 &       0.015 &    0.495 &     1.027                                                    \\
TSHestonMR                     &                   0.023 &                  0.048 &               0.031 &             0.085 &         0.476 &          0.741 &         0.266 &     0.484 &      1.163                                                    \\
CSHestonMOM                     & -0.000 &                 0.015 &               0.010 &            0.031 &       -0.005 &        -0.007 &       -0.002 &    0.507 &     0.973                                                   \\
CSHestonMR                     &                   0.009 &                  \textbf{0.015*} &               \textbf{0.010*} &             \textbf{0.028*} &         0.565 &          0.862 &         0.301 &     0.500 &      1.109                                                    \\

\midrule
{\underline{\textbf{Deep Learning Models}}}        &                    &                 &                                                                           &                 &                 &                  &                 &                                                                             &                                                          \\
Linear                    & 0.021 &                 0.018 &              0.014 &            0.036 &        1.258 &         1.672 &        0.754 &    0.561 &     1.073                                                    \\

MLP              & 0.016 &                 0.017 &              0.013 &            0.034 &        0.954 &          1.262 &        0.504 &    \textbf{0.580*} &     0.939                                                    \\
CNN                       & 0.009 &                 0.016 &              0.012 &            0.035 &       0.534 &        0.717 &       0.240 &    0.507 &     1.199                                                    \\
LSTM                    & 0.026 &                 0.019 &               0.014 &            0.031 &        \textbf{1.399*} &         \textbf{1.917*} &        \textbf{0.850*} &     0.557 &     1.231                                                    \\ 
\bottomrule \hline

\end{tabular}
}
\begin{flushleft}$_{\text{\ \ \ \ (bold and * denotes best performing strategy for each column)}}$\end{flushleft}
\vfill
\end{table*}

\begin{table*}[htbp]
\centering
\caption{Performance Metrics -- Rescaled to Target Volatility}
\label{table:scaled_signals}
\resizebox{\textwidth}{!}{
\begin{tabular}{llllllllll}
\hline \toprule
& \textbf{E[Return]} & \textbf{Vol.} & \textbf{\begin{tabular}[c]{@{}l@{}} Downside \\ Deviation \end{tabular}} & \textbf{MDD} & \textbf{Sharpe} & \textbf{Sortino} & \textbf{Calmar} & \textbf{\begin{tabular}[c]{@{}l@{}} Hit \\ Rate \end{tabular}} & \textbf{$\mathbf{\frac{\text{Ave. P}}{\text{Ave. L}}}$} \\ 
\midrule
{\underline{\textbf{Benchmarks}}} &   &   &   &   &   &   &   &   &  \\
Long Only                 &                   0.112 &                  0.160 &               0.088 &             0.292 &         0.697 &          1.275 &         0.383 &     0.451 &      1.384                                                    \\
TSMR                      &                   0.123 &                  0.161 &               \textbf{0.087*} &             0.254 &         0.762 &          1.414 &         0.484 &     0.445 & \textbf{1.441*}                                                    \\ 
MACDMR                      &                   0.106 &                  0.162 &               0.088 &             0.264 &         0.655 &          1.207 &         0.402 &     0.443 &      1.424                                                    \\ 
TSHestonMR                    &                   0.098 &                  \textbf{0.157*} &               0.100 &             \textbf{0.191*} &         0.626 &          0.984 &         0.514 &     0.484 &      1.193                                                    \\
CSHestonMR                     &                   0.091 &                  0.159 &               0.104 &             0.318 &         0.573 &          0.878 &         0.286 &     0.500 &      1.112                                                    \\

\midrule
{\underline{\textbf{Deep Learning Models}}}        &                    &                 &                                                                           &                 &                 &                  &                 &                                                                             &                                                          \\
Linear                    & 0.245 &                 0.190 &              0.146 &            0.276 &        1.290 &         1.690 &        0.917 &    0.561 &     1.094                                                    \\

MLP              & 0.167 &                 0.186 &              0.141 &            0.344 &        0.895 &          1.199 &        0.523 &     \textbf{0.580*} &     0.929                                                    \\
CNN                       & 0.119 &                 0.217 &              0.163 &            0.435 &       0.551 &        0.736 &       0.284 &    0.507 &     1.216                                                    \\
LSTM                    & \textbf{0.266*} &                 0.200 &               0.144 &            0.274 &         \textbf{1.329*} &          \textbf{1.862*} &         \textbf{0.974*} &     0.557 &     1.212                                                    \\ 
\bottomrule \hline

\end{tabular}
}
\end{table*}

\begin{figure*}[htbp]
\centering
\includegraphics[width=0.74\linewidth]{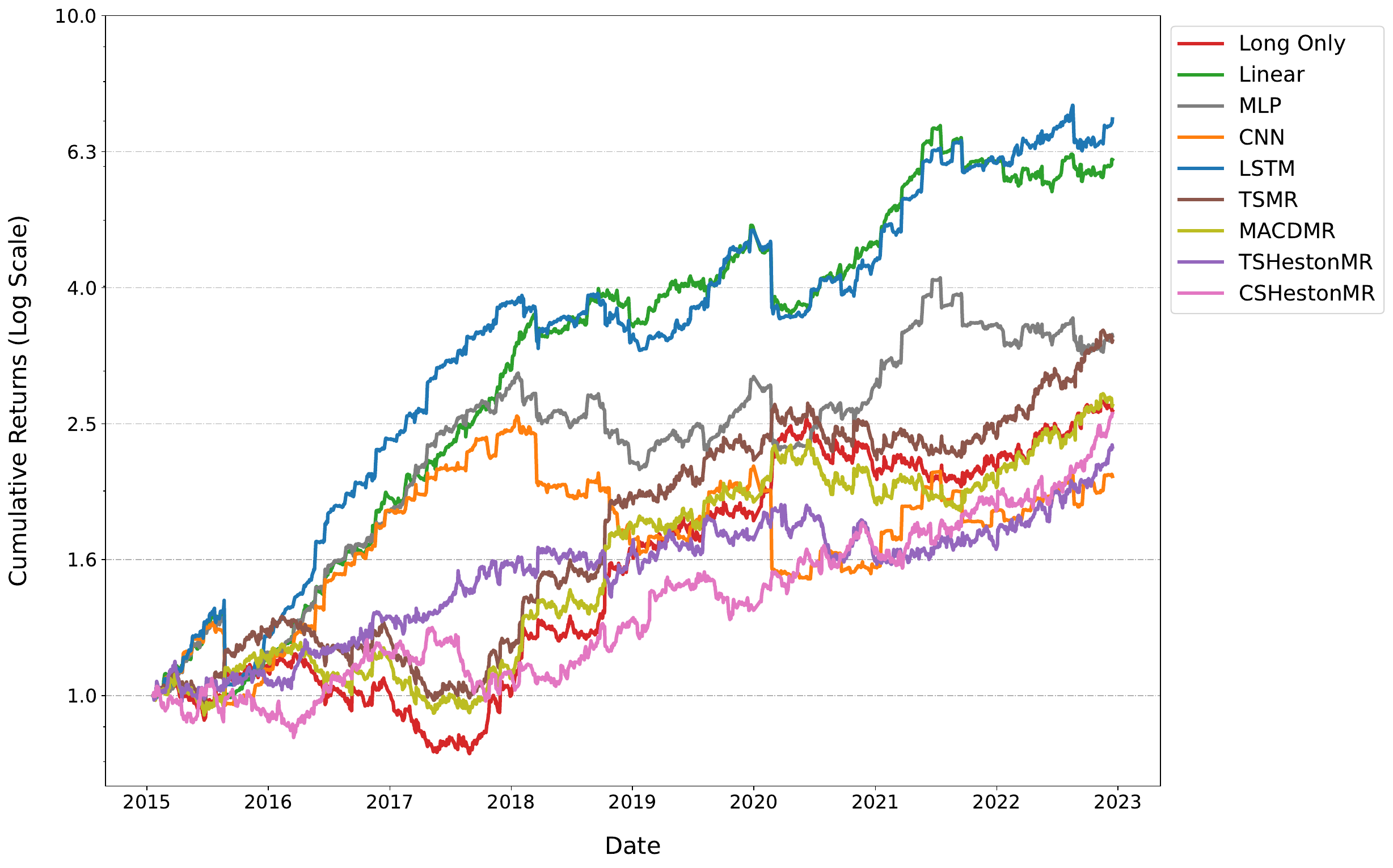} 
\vspace{-0.3cm}
\caption{Cumulative Returns - Rescaled to Target Volatility}
\label{fig:cumulative_returns_scaled_signals}
\Description[Plot showing Cumulative Returns - Rescaled to Target Volatility]{Plot showing Cumulative Returns - Rescaled to Target Volatility, LSTM and Linear models exhibit highest cumulative returns}
\end{figure*}

\begin{table*}[htbp]
\centering
\caption{Impact of Transactions Costs on Sharpe Ratio -- Rescaled to Target Volatility}
\label{table:sharpe_ratio_and_transaction_cost}

\begin{tabular}{lllllllll}
\hline \toprule
\textbf{Transaction Costs (bps)} & \textbf{0.0} & \textbf{1.0} & \textbf{2.0} & \textbf{3.0} & \textbf{5.0} & \textbf{10.0} & \textbf{20.0} & \textbf{50.0} \\ 
\midrule

{\underline{\textbf{Benchmarks}}}        &                    &                 &                                                                           &                 &                 &                  &                 &                                                                                                                                       \\
Long Only                    &  0.697 &  0.695 &  0.692 &  0.690 &  0.685 &  0.673 &  0.650 &  0.578                                                   \\
TSMR                    &  0.762 &  0.757 &  0.752 &  0.747 &  0.737 &  0.713 &  0.663 &  0.514                                                   \\
MACDMR                    &  0.655 &  0.646 &  0.638 &  0.629 &  0.612 &  0.568 &  0.481 &  0.219                                                   \\
TSHestonMR                    &  0.626 &  0.621 &  0.615 &  0.610 &  0.600 &  0.574 &  0.522 &  0.367                                                   \\
CSHestonMR                    &  0.573 &  0.570 &  0.567 &  0.563 &  0.557 &  0.541 &  0.508 &  0.410                                                   \\
\midrule

{\underline{\textbf{Deep Learning Models}}}        &                    &                 &                                                                           &                 &                 &                  &                 &                                                                                                                                       \\
LSTM                    & \textbf{1.329*} &  \textbf{1.310*} &  \textbf{1.291*} &  \textbf{1.272*} &  \textbf{1.235*} &  1.140 &  0.952 &  0.388                                                   \\
LSTM + TC Reg.            & 1.282 &  1.270 &  1.259 &  1.247 &  1.223 &  \textbf{1.164*} &  \textbf{1.045*} &  \textbf{0.689*}                                                   \\
\bottomrule \hline
\end{tabular}

\vfill
\end{table*}

We use the following annualized metrics to evaluate the out-of-sample performance of all strategies:
\begin{enumerate}[label = \textbf{\Roman*.}]
    \item \textbf{Profitability Measures} -- Expected Returns ($\mathbb{E}[\text{Returns}]$), Hit Rate
    \item \textbf{Risk Measures} -- Volatility (Vol.), Downside Deviation, Maximum Drawdown (MDD)
    \item \textbf{Performance Ratios} -- Sharpe, Sortino and Calmar Ratios, Average Profit over Loss $\left(\frac{\text{Ave. P}}{\text{Ave. L}} \right)$
\end{enumerate}

We present the aggregated out-of-sample performance metrics of all strategies computed using the overall returns according to Equation (\ref{eqn:strategy}). Firstly, we present the performance of all strategies from their raw signal outputs in Table \ref{table:raw_signals}. We then apply to all strategies (excluding unprofitable strategies) an additional layer of volatility scaling to target an annualized volatility of $15\%$ at the portfolio level and report the performance in Table \ref{table:scaled_signals} and plot their cumulative returns in Figure \ref{fig:cumulative_returns_scaled_signals}. This adjustment at the portfolio level facilitates comparison between individual strategy returns in line with our 15\% volatility target. For each Heston portfolio (TSHestonMOM, TSHestonMR, CSHestonMOM, CSHestonMR), we report results for the best performing lookback period for the sake of brevity. In this section, we report the performance of all strategies without factoring in transaction costs to evaluate their raw predictive ability. In Section \ref{section:transaction_cost_impact}, we include an analysis of the impact of transaction costs and the effect of turnover regularization.

From Table \ref{table:raw_signals}, we find that the Long Only straddle portfolio was a profitable strategy over the backtest period. We find this observation interesting, running contrary to \cite{hu2023profits} who show that retail and institutional investors executing short volatility strategies tend to perform well. Furthermore, a Long Only options portfolio would typically benefit from limited downside exposures as opposed to Short Only. Turning our attention to trend-based strategies, we observe that mean-reversion portfolios (TSMR, MACDMR, TSHestonMR, CSHestonMR) exhibit positive performances compared to their opposite momentum counterparts (TSMOM, MACD, TSHestonMOM, CSHestonMOM), which were generally unprofitable over the backtest period. We note that both TSMR and CSHestonMR exhibited only slight performance improvements over Long Only. We obtain similar results for the Heston portfolios as reported in \cite{heston2023option}, who document significant reversals in option returns at short-term horizons, and our best performing Heston models correspond mostly with strategies adopting the shortest $n = 1$ month lookback period. Moving on to our deep learning models, apart from the CNN, we observe that the Linear, MLP and LSTM exhibit a clear disparity in performance above all other strategies as seen from their higher performance ratios.

Referring to Table \ref{table:scaled_signals}, with the implementation of volatility targeting at the portfolio level, we observe modest improvements in performance across all profitable benchmarks. Portfolio volatility targeting resulted in minimal changes to the deep learning models, allowing them to retain a large gap in their performance ratios over the benchmarks. In particular, we see that the Linear and LSTM models exhibited the best performances, outperforming the benchmarks by roughly twice in their Sharpe ratios of 1.290 and 1.329 respectively. We note that the simplest Linear model was able to perform the MLP, most likely due to our introduction of an L1 regularization penalty only for the Linear model during training. Examining the period during the COVID-19 market selloff, we see clearly from Figure \ref{fig:cumulative_returns_scaled_signals} that Long Only exhibited sharp gains during the COVID-19 market selloff at the start of 2020, demonstrating the profitability of long straddles during periods of high market volatility. While we observe that the deep learning models experienced brief drawdowns during the selloff, performance of the models swiftly recovered following the market rebound.

\subsection{Transaction Costs and Turnover Regularization}
\label{section:transaction_cost_impact}

Some of the key challenges of rebalancing an options portfolio include market microstructure considerations arising from market liquidity and bid-ask spreads, which can result in high transaction costs. To examine the impact of transaction costs on the profitability of all strategies, we first compute Sharpe ratios adjusted for transaction costs by taking into account turnover-adjusted returns: 
\begin{equation}
\label{eqn:turnover_adjusted_returns}
\tilde{r}_{t,t+1}^{\text{STRATEGY}} = \frac{1}{N_t} \sum_{i=1}^{N_t} \bigg(R_i(t) -  c \cdot \sigma_{\text{tgt}} \left| \frac{X_t^{(i, \text{straddle})}}{\sigma_t^{(i, \text{straddle})}} -  \frac{X_{t-1}^{(i, \text{straddle})}}{\sigma_{t-1}^{(i, \text{straddle})}} \right| \bigg)
\end{equation}
where $c$ represents a measure of average transaction costs in basis points. Focusing on the LSTM model, we observe from Table \ref{table:sharpe_ratio_and_transaction_cost} that the model maintains superior risk-adjusted performance over the best performing benchmark - TSMR up to transaction costs of $c = 20$ bps, deteriorating at higher transaction costs of $c = 50$ bps.

Following the methodology detailed in \cite{lim2019enhancing, tan2023spatio}, we modify the training loss to utilize turnover-adjusted returns as defined in Equation (\ref{eqn:turnover_adjusted_returns}), which in effect results in optimizing the Sharpe ratio while regularizing for the turnover generated by the trading signals of the LSTM. Based on Table \ref{table:sharpe_ratio_and_transaction_cost}, we see that turnover regularization further enhances the performance of the LSTM for high transaction costs of $c = 10$ to $50$ bps, allowing the regularized model to outperform other strategies at prohibitively high levels of transaction costs.

\section{Conclusions}
\label{conclusions}
We present a general end-to-end framework for trading options using a highly data-driven machine learning algorithm, adopting the point of view of an active investor seeking to profit from options trading. Departing from conventional approaches that typically rely on specific market dynamics or option pricing models, we train end-to-end neural networks to directly learn mappings from options data to optimal trading positions, removing the need to simulate market processes, price options or predict option returns. Backtesting our approach on portfolios of delta-neutral equity options, our models demonstrate significant improvements in risk-adjusted performance when calibrated using the Sharpe ratio. Crucially, our framework is agnostic to specific underlying market assumptions, potentially allowing for further extensions to a broader set of derivatives or complex instruments where data is available.

\begin{acks}
We would like to thank the Oxford-Man Institute of Quantitative Finance for providing compute resources. Wee Ling Tan thanks Bryan Lim, Martin Luk, Guillaume Andrieux and Hans Buehler for their insightful comments.
\end{acks}

\bibliographystyle{ACM-Reference-Format}
\bibliography{bibliography}

\appendix

\section{Hyperparameter Optimization}
\label{appendix_a}
\begin{table}[H]
\centering
\caption{Hyperparameter Search Range}
\label{table:hyperparameter_search}
\begin{tabular}{lll}
\toprule
\textbf{Hyperparameters}           & \textbf{Search Grid}              \\ \midrule
Minibatch Size                     & 32, 64, 128, 256             \\
Dropout Rate                       & 0.1, 0.2, 0.3, 0.4, 0.5           \\
Hidden Layer Size                  & 5, 10, 20, 40, 80, 160       \\
Learning Rate                      & $10^{-5},~ 10^{-4},~ 10^{-3},~ 10^{-2},~ 10^{-1},~ 10^{0}$                       \\
Max Gradient Norm                  & $10^{-4},~ 10^{-3},~ 10^{-2},~ 10^{-1},~ 10^{0},~ 10^{1}$                       \\
\bottomrule
\end{tabular}
\end{table}

\end{document}